\documentclass[prd,showpacs,amsmath,amssymb,floatfix]{revtex4}
\usepackage{graphicx,color,dcolumn,,booktabs}
\begin{document}
\title{Discovery potential for charmonium-like state $Y(3940)$ by the meson photoproduction}
\author{Jun He$^{1,2}$}
\email{junhe@impcas.ac.cn} \author{Xiang Liu$^{1,3}$\footnote{Corresponding author}}\email{xiangliu@lzu.edu.cn}
\affiliation{$^1$Research Center for Hadron and CSR Physics,
Lanzhou University $\&$ Institute of Modern Physics of CAS, Lanzhou 730000, China
\\$^2$Institute of Modern Physics, Chinese Academy of Sciences, Lanzhou 730000, China\\
$^3$School of Physical Science and Technology, Lanzhou University, Lanzhou 730000, China}
\date{\today}
\begin{abstract}

In this work, we investigate the discovery potential for $Y(3940)$ by the photoproduction process
$\gamma p\to Y(3940)p$. The numerical result shows that the upper (lower) limit
of the total cross section for $\gamma p\to Y(3940)p$ is up to the order of
1 $nb$ (0.1 $\mu b$). Additionally, the background analysis and the Dalitz plot
relevant to the production of $Y(3940)$ are studied. The Dalitz plot analysis of $Y(3940)$ production indicates
that $Y(3940)$ signal can be distinguished from the background clearly. The lower limit
of the number of events of $Y(3940)$ reaches up to 10/0.02GeV$^2$ for $1\times 10^9$ collisions of $\gamma p$ by studying the invariant mass spectrum of $J/\psi\omega$. An experimental search for $Y(3940)$ by the meson photoproduction
is suggested.

\end{abstract}

\maketitle
\section{introduction}\label{sec1}

With the development of the experimental technology,
experiments have announced a series of charmonium-like states $X,\,Y,\,Z$ in the past six years. Among
these charmonium-like states, $Y(3940)$ is an enhancement near the threshold of
$J/\psi\omega$ in exclusive $B\to K\omega J/\psi$ decay
\cite{Abe:2004zs,Aubert:2007vj}. The experimental information given by the Belle
and Babar Collaborations is summarized in Table \ref{mass-width}.
\begin{table}[htb] \begin{tabular}{cccccc}\toprule[1pt] Experiment&Mass
	(MeV)&Width (MeV)&$\mathcal{B}[B\to K Y(3940)]\mathcal{B}[Y(3940)\to
	J/\psi \omega]$\\\midrule[1pt] Belle
	\cite{Abe:2004zs}&$3943\pm11(\mathrm{stat})\pm
	13(\mathrm{syst})$&$87\pm22(\mathrm{stat})\pm
	26(\mathrm{syst})$&$(7.1\pm
	1.3(\mathrm{stat})\pm3.1(\mathrm{syst}))\times10^{-5}$\\ Babar
	\cite{Aubert:2007vj}&$3914^{+3.8}_{-3.4}(\mathrm{stat})\pm
	2.0(\mathrm{syst})$&$34^{+12}_{-8}(\mathrm{stat})\pm
	5(\mathrm{syst})$&$(4.9^{+1.0}_{-0.9}(\mathrm{stat})\pm
	0.5(\mathrm{syst}))\times10^{-5}$\\\bottomrule[1pt]
\end{tabular}\caption{The mass, width and product branching fraction of
$Y(3940)$ measured by the Belle and Babar experiments.\label{mass-width}}
\end{table}
Although the values of mass and width given by the Babar
Collaboration are smaller than those reported by the Belle Collaboration, the
measured product branching ratios by Belle and Babar agree with each other.  As
indicated in Refs. \cite{Zhu:2007wz,Godfrey:2008nc}, taking the typical value
for allowed $B\to K+charmonium$ decays ($\mathcal{B}[B\to KY(3940)]\leq
1\times10^{-3}$), one obtains the lower limit of the decay width of $Y(3940)\to
J/\psi\omega $, which means $\Gamma[Y(3940)\to J/\psi \omega]\geq 1$ MeV and
$\Gamma[Y(3940)\to J/\psi \omega]\geq 4$ MeV corresponding to Babar and Belle
results, respectively. Thus, to some extent, explaining $Y(3940)$ as
a conventional charmonium state is problematic due to the unusual decay width of
$Y(3940)\to J/\psi \omega$ \cite{Godfrey:2008nc}.

In the past years, theorists have been puzzled with
the underlying structure of $Y(3940)$  after the observation of $Y(3940)$. With the observation of $Y(4140)$ by the
CDF \cite{Aaltonen:2009tz}, the similarities between $Y(4140)$ and $Y(3940)$ provide the hint to
reveal the structure of $Y(3940)$. Thus, the explanations of $D^*\bar{D}^*$ and
$D_{s}^*\bar{D}_s^*$ respectively corresponding to $Y(3940)$ and $Y(4140)$ were
proposed in Refs. \cite{Liu:2009ei,Branz:2009yt}. The possible quantum number
assignment includes $J^{PC}=0^{++}$ or $J^{PC}=2^{++}$ \cite{Liu:2009ei}.

It is well-known that most of charmonium-like states are only observed in the
two $B$ factories by $B$ meson decay, which provides the $c\bar{c}$ rich
environment. However, searching for $X,\,Y,\,Z$ charmonium-like states by other
production processes is an interesting topic, which will not only confirm these
observed charmonium-like states, but also be helpful to study their
underlying structures. For example, theorists carried out the relevant studies of
the production of $Z^\pm(4430)$, which is an enhancement in the $\psi^\prime
\pi^\pm$ invariant mass spectrum \cite{:2007wga}. Liu, Zhao, and Close once proposed to search for
$Z^\pm(4430)$ by the meson photoproduction process \cite{Liu:2008qx}.
Additionally, the authors in Ref. \cite{Ke:2008kf} suggested an experiment to find
the signal of $Z^\pm(4430)$ by the  nucleon-antinucleon scattering.

Meson photoproduction is an important way to study hadron spectroscopy,
such as the baryon resonances and charmonium with a higher beam
energy \cite{Chekanov:2002xi,Chekanov:2002at,Chekanov:2004mw}. The thresholds for $X,Y,Z$ photoproductions fall
in the energy region of the HERA experiment. Thus, it is interesting to study whether
there exists a potential to find $Y(3940)$ by meson photoproduction. In
this work we will be dedicated to exploring $Y(3940)$ production by the
photoproduction process.

The paper is organized as follows. After the introduction, we present the
calculation of $Y(3940)$ photoproduction. In Sec. \ref{Sec:Background},
the possible background relevant to the production of $Y(3940)$ is discussed
and the Dalitz plot is presented. The last section is the conclusion and
discussion.

\section{$Y(3940)$ in the photoproduction process}\label{Sec:Y production}

\begin{center}
\begin{figure}[htb]
\begin{tabular}{cc}
\scalebox{0.8}{\includegraphics{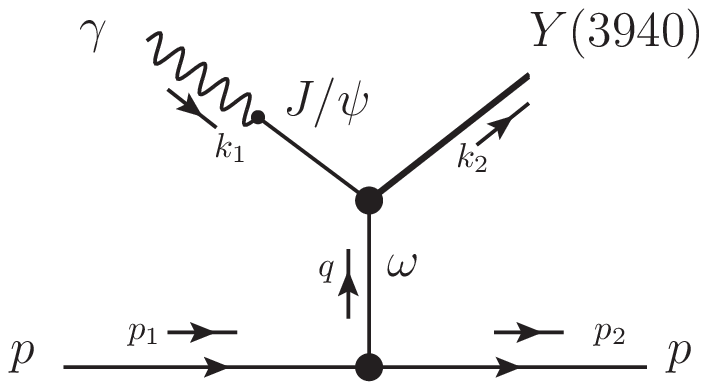}}&\scalebox{0.8}{\includegraphics{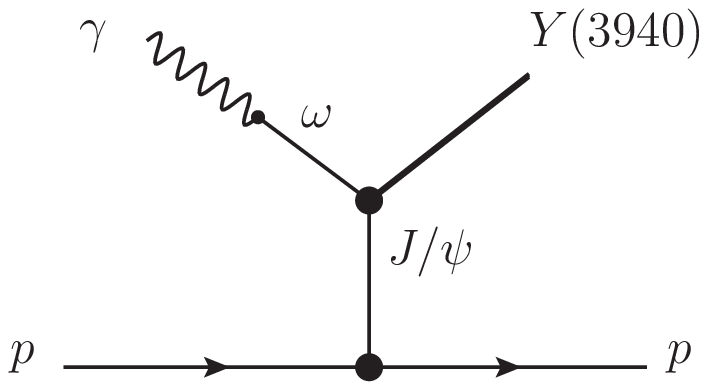}}\\
(a)&(b)\\
\end{tabular}
\caption{The $Y(3940)$ photoproductions through $\omega$ and $J/\psi$ exchanges. \label{Fig:gpept}}
\end{figure}
\end{center}

In this section, we will discuss the production probability of $Y(3940)$
through calculating the cross section of the photoproduction process.  As shown in
the introduction, Babar and Belle reported the observation of $Y(3940)$ in the
decay channel to $J/\psi$ and $\omega$. Since both $J/\psi$ and $\omega$
interact with a photon via the well-known vector meson dominance mechanism, the
$Y(3940)$ photoproduction process can be described by Fig. \ref{Fig:gpept}. Here,
the $t$ channel is the dominant process due to the threshold of $Y(3940)$ photoproduction
being about 4 GeV.

Considering the coupling between $J/\psi$ and the nucleon being obviously weaker than that
between $\omega$ and the nucleon, in this work one mainly discusses the
photoproduction process depicted in Fig. \ref{Fig:gpept} (a). Here, the $\omega$
meson is the exchanged state and $J/\psi$ is as the intermediate vector meson
coupling to the incoming photon.  Since the contribution from the pomeron exchange is negligible 
as discussed in Ref. \cite{Liu:2008qx}, we do not consider the
contribution of the pomeron exchange to the $Y(3940)$ photoproduction process in this work .

As discussed in Ref.~\cite{Liu:2009ei}, the charmonium-like state $Y(3940)$ may
have a quantum number $J^{PC}=0^{++}$ or $2^{++}$. The vertices depicting the interaction of
$Y(3940)$ and $J/\psi\omega$ are
\cite{Black:2002ek,Renner:1971sj,Oh:2003aw,Lansberg:2009xh}

\begin{eqnarray}
	\langle J/\psi(k_1) \omega(k_2)| Y(3940)\rangle&=&\frac{g_{Y J/\psi\omega}}{M_Y}\epsilon_{1}^\mu\epsilon_{2}^\nu~T_{\mu\nu},  \quad \quad \quad \quad \,\,\,\,\mathrm{for} \quad \quad 0^{++}\label{1}\label{lag1}\\
	\langle J/\psi(k_1) \omega(k_2)| Y(3940)\rangle&=&\frac{g_{Y J/\psi\omega}}{M_Y}\epsilon_1^\mu\epsilon_2^\nu \epsilon^{\alpha\beta} M_{\mu\nu,\alpha\beta},\quad \quad \mathrm{for} \quad \quad 2^{++}\label{2}\label{lag2}
\end{eqnarray}
with
\begin{eqnarray}
	T^{\mu\nu}_{k_1,k_2}&=&g^{\mu\nu}k_1\cdot k_2-k_1^\nu k_2^\mu,\\
	M^{\mu\nu,\alpha\beta}_{k_1,k_2}&=&g^{\nu\beta} k_1^\alpha k_2^\mu
	+ g^{\mu\alpha}k_1^\nu k_2^\beta-g^{\mu\nu}k_1^\alpha k_2^\beta
	-g^{\mu\alpha}g^{\nu\beta}k_1\cdot k_2,
\end{eqnarray}
where $\epsilon^\mu$ and  $\epsilon^{\mu\nu}$ are the polarization vector and
the polarization tensor corresponding to $Y(3940)$ with $J^P=0^{++}$ and
$J^P=2^{++}$, respectively. The Lagrangians listed in Eqs. (\ref{lag1}) and
(\ref{lag2}) are derived using gauge invariance and vector and tensor meson dominance.
We adopt the lower (upper) limit of the decay
width of $Y(3940)\to J/\psi \omega$, $\Gamma[Y(3940)\to J/\psi \omega]= 1(34)$
MeV \cite{Aubert:2007vj,Zhu:2007wz,Godfrey:2008nc}, to determine the lower (upper) limit of the coupling constant: $g_{Y
J/\psi\omega}=0.264 (1.54)$ GeV$^{-1}$ for scalar $Y(3940)$ or $g_{Y
J/\psi\omega}=0.461 (2.69)$ GeV$^{-1}$ for tensor $Y(3940)$. Additionally, for
describing the vertex with the off-shell $\omega$, we introduce the monopole
form factor
${F}_{YJ/\psi\omega}(q^2)=(\Lambda_Y^2-m_\omega^2)/(\Lambda_Y^2-q^2)$, where we
set the cutoff $\Lambda_Y$ as the mass of  $J/\psi$ as suggested in Ref.
\cite{Liu:2008qx}.

In the vector meson dominance mechanism, the Lagrangian depicting the coupling of the intermediate state $J/\psi$ with photon is written as
\begin{eqnarray}
	{\cal L}_{{J/\psi}\gamma}=-\frac{eM_{J/\psi}^2}{f_{J/\psi}}V_\mu A^\mu,
\end{eqnarray}
where $M_{J/\psi}$ and $f_{J/\psi}$ denote the mass and the decay constant of $J/\psi$ respectively.
In terms of the decay width of $J/\psi\to e^+e^-$ \cite{Amsler:2008zzb} $$\Gamma_{J/\psi\to e^+e^-}= 5.55\pm0.14\pm0.02\quad \mathrm{keV},$$ one obtains the parameter
$e/f_{J/\psi}=0.027$.

We adopt the effective Lagrangian
\begin{eqnarray}
	{\cal L}_{\omega NN}=-g_{\omega NN}\bar{N}(\gamma\cdot\omega-\frac{\kappa_\omega}{2M_N}\sigma^{\mu\nu}\partial_\nu\omega_\mu)N
\end{eqnarray}
to describe the coupling between $\omega$ and the nucleon. Besides, one introduces the monopole form factor $F_{\omega NN}=({\Lambda_{\omega}^2-m_\omega^2})/({\Lambda_{\omega}^2-q^2})$ in $NN\omega$ vertex
to compensate the off-shell effect of the $\omega$ meson.
Here, $g_{\omega NN}=12$, $\kappa_\omega=0$
and $\Lambda_{\omega}=1.2$ GeV \cite{Gasparyan:2003fp}. The vertex of $J/\psi\omega Y$ is result of the vector meson dominance mechanism. Thus,
there exists the difference of the vertex of $NN\omega$ from the vertex of $J/\psi\omega Y$.
We distinguish the vertex of $NN\omega$ and the vertex of $J/\psi\omega Y$
by introducing different cutoff $\Lambda_Y$ and $\Lambda_\omega$ corresponding to the vertex of $NN\omega$ and the vertex of $J/\psi\omega Y$ respectively.

The differential cross section for $Y(3940)$ photoproduction shown in Fig. \ref{Fig:gpept} (a) reads as
\begin{eqnarray}
	\frac{d\sigma}{dt}=\frac{1}{256\pi s}\frac{1}{|k_{1cm}|^2}
	\sum_{pol}|{\cal T}_{fi}|^2,
\end{eqnarray}
where $s=(p_1+k_1)^2=(p_2+k_2)^2=W^2$ and $t=(k_2-k_1)^2=(p_1-p_2)^2=q^2$ denote the Mandelstam variables. $k_{1cm}$ is the photon energy
in the center of mass frame of $\gamma p$ scattering.

The amplitude ${\cal T}_{fi}$ for the production of scalar $Y(3940)$ is obtained
\begin{eqnarray}
	{\cal T}_{fi}&=&\Big(g_{\omega NN}\frac{g_{YJ/\psi\omega}}{M_Y}\frac{e}{f_{J/\psi}}\Big)\bar{u}(p_2)O^\alpha u(p_1)
	~\epsilon_{1\alpha},
\end{eqnarray}
where
$$O^{\alpha}=G^\mu~\frac{\bar{g}_{\mu\nu}}
{q^2-m_\omega^2} A^{\nu\alpha},\quad G^\mu=\gamma^\mu {F}_{\omega NN}(q^2),\quad A^{\nu\alpha}=T_{k_1,q_1}^{\alpha\nu}~{F}_{YJ/\psi\omega}(q^2).$$
Analogously, for the $\gamma p\to Y(3940)p$ with tensor $Y(3940)$, the amplitude ${\cal T}_{fi}$ is
\begin{eqnarray}
	{\cal T}_{fi}&=&\Big(g_{\omega NN}\frac{g_{YJ/\psi\omega}}{M_Y}\frac{e}{f_{J/\psi}}\Big)\bar{u}(p_2)O^{\alpha,\beta\gamma} u(p_1)
	~\epsilon_{1\alpha}~\epsilon_{\beta\gamma},
\end{eqnarray}
with $$O^{\alpha,\beta\gamma}=G^\mu~\frac{\bar{g}_{\mu\nu}}
{q^2-m_\omega^2} A^{\nu\alpha,\beta\gamma},\quad G^{\mu}=\gamma^\mu {F}_{\omega NN}(q^2),\quad A^{\nu\alpha,\beta\gamma}=M_{k_1,q_1}^{\alpha\nu,\beta\gamma}{F}_{YJ/\psi\omega}(q^2).$$
In the above expressions, we define $\bar{g}_{\mu\nu}=-g_{\mu\nu}+q_\mu q_\nu/m^2_\omega$. Since, the longitudinal part of the omega-meson
propagator does not contribute to the amplitudes, we use $\bar{g}_{\mu\nu}=-g_{\mu\nu}$ instead of $\bar{g}_{\mu\nu}=-g_{\mu\nu}+q_\mu q_\nu/m^2_\omega$ for writing out the amplitude.

\begin{center}
\begin{figure}[htb]
\begin{tabular}{cc}
\scalebox{0.8}{\includegraphics{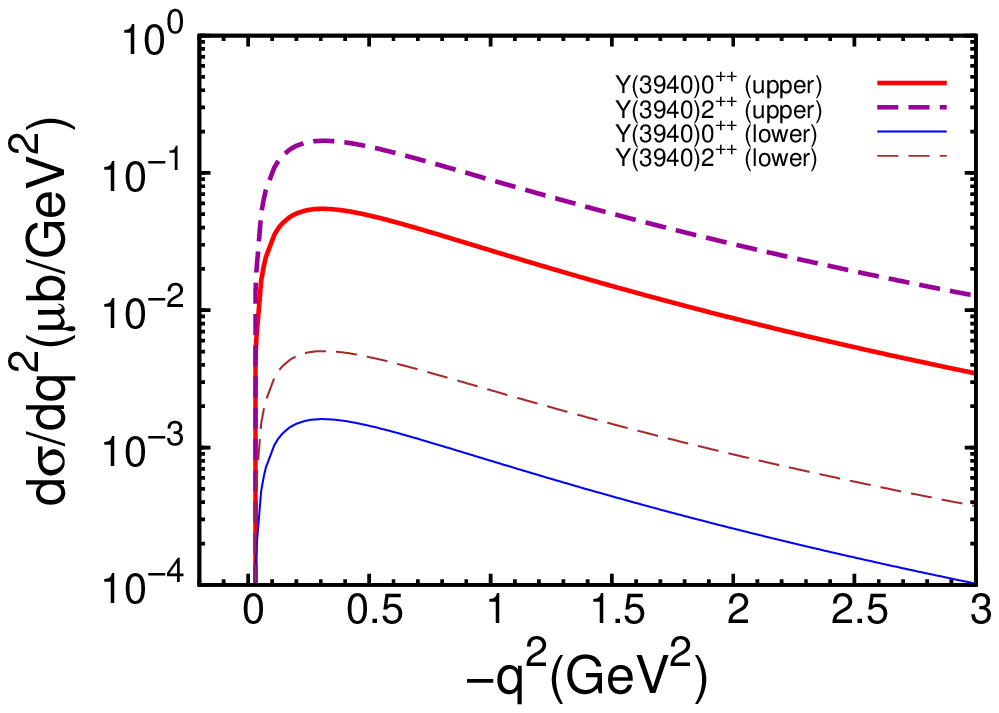}}&\scalebox{0.8}{\includegraphics{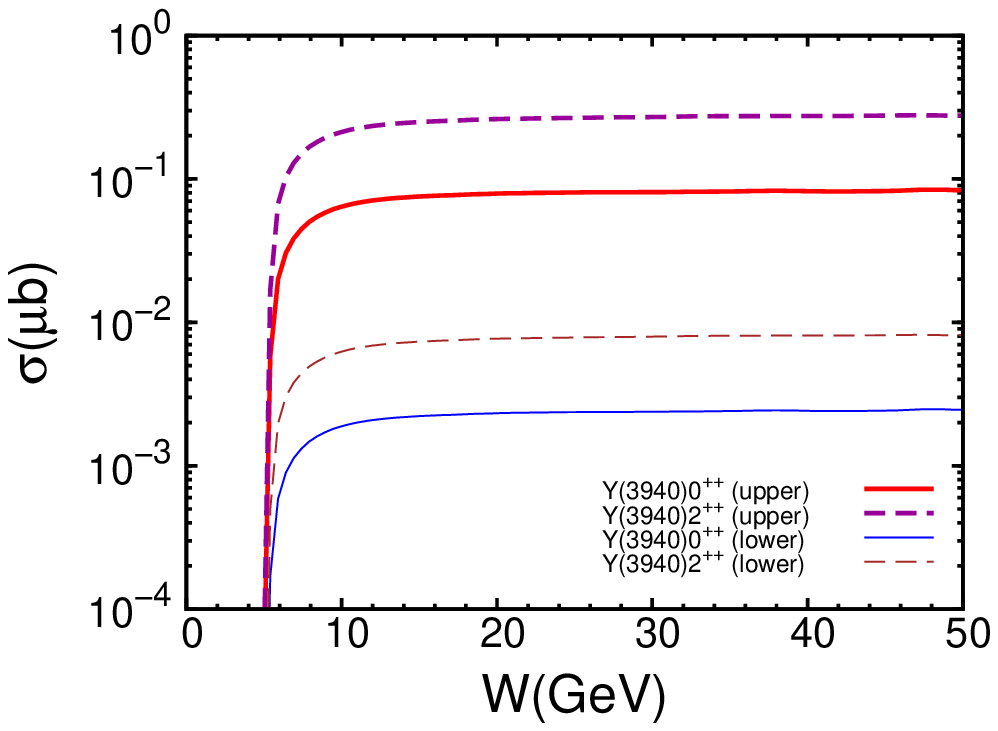}}\\
(a)&(b)\\
\end{tabular}
\caption{The upper and lower limits of the cross sections for the production of $Y(3940)$
with quantum number $0^{++}$ or $2^{++}$. (a) The variation of the differential cross sections
of $\gamma p\to Y(3940)p$ to $-q^2$. (b) The dependence of the total cross section of $\gamma p\to Y(3940)p$ on the total energy in the center of mass frame of $\gamma p$ scattering.
 \label{Fig:gNYNTCS}}
\end{figure}
\end{center}

With the above preparation, we obtain the differential cross section for
$Y(3940)$ production, which is shown in Fig. \ref{Fig:gNYNTCS}. After
integrating over the range of $t$, the total cross section
$\sigma(\gamma p\to Y(3940)p)$ is presented in Fig. \ref{Fig:gNYNTCS}.

These results show that a peak appears at the low $-q^2$ region. The line shape
of the total cross section goes up very rapidly near the threshold, while the
line shape of the differential cross section becomes flat with the increase of
$-q^2$.
The total cross section for the production of tensor $Y(3940)$ is about 5
times larger than that for the production of the scalar $Y(3940)$.
The total cross section is proportional to the square of the coupling
$g_{YJ/\psi\omega}$, which indicates that the cross section is also
proportional to the decay width of $Y(3940)\to J/\psi\omega$. Since the
concrete value of the decay width of $Y(3940)\to J/\psi\omega$ is undetermined
by theory and experiment, we only give the upper and lower limits of
$\sigma(\gamma p\to Y(3940)p)$ in this work according to the upper and lower limits of the decay width of $Y(3940)\to J/\psi\omega$. The upper limit of $\sigma(\gamma p\to
Y(3940)p)$ under the two assignments of the quantum number for $Y(3930)$ is on
the order of 0.1 $\mu b$, which is comparable with the cross section of
$J/\psi$ photoproduction in the HERA experiment
\cite{Chekanov:2002xi,Chekanov:2004mw}.  The lower limit of $\sigma(\gamma p\to
Y(3940)p)$ is a few $nb$, which is comparable with the cross section of
$\psi(2S)$ \cite{Chekanov:2002at} or bottomonium $\Upsilon$ photoproduction in
HERA \cite{Chekanov:2009xb,Breitweg:1998ki}.

\section{Background analysis and Dalitz plot}\label{Sec:Background}

Since the produced $Y(3930)$ decays into $J/\psi \omega$, the experimental
channel relevant to $Y(3940)$ is  $\gamma p\to J/\psi\omega p$.  As indicated
in our calculation, the total cross section for tensor $Y(3940)$ production is much
larger than that for the scaler one. In the following, with the scalar $Y(3940)$
photoproduction as an example, we discuss the background analysis relevant to
$Y(3940)$ photoproduction.

\begin{center}
\begin{figure}[htb]
\begin{tabular}{cc}
\scalebox{0.8}{\includegraphics{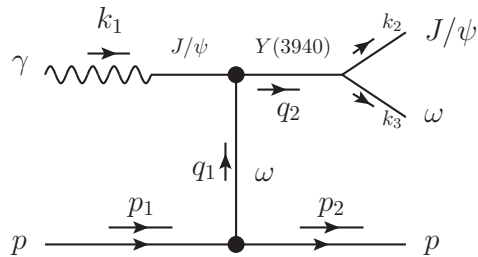}}
\end{tabular}
\caption{The diagram depicting $\gamma p\to Y(3940)p\to J/\psi\omega p$ through $\omega$ exchange.\label{Fig:gppop}}
\end{figure}
\end{center}

First we consider the $\gamma p\to J/\psi\omega p$ process as shown in Fig.
\ref{Fig:gppop}, which corresponds to $Y(3940)$ production.  The amplitude for
$\gamma p\to Y(3940)p\to J/\psi\omega p$ depicted in Fig. \ref{Fig:gppop} can
be written as

\begin{eqnarray}
	{\cal T}_{fi}&=&\Big[g_{\omega NN}\Big(\frac{g_{YJ/\psi\omega}}{M_Y}\Big)^2\frac{e}{f_{J/\psi}}\Big]
\bar{u}(p_2)O^{\alpha\beta\gamma}u(p_1)
\epsilon_{1\alpha}~\epsilon_{2\beta}\epsilon_{3\gamma}
\end{eqnarray}
with \begin{eqnarray}O^{\alpha\beta\gamma}&=&G^\mu~\frac{\bar{g}_{\mu\nu}}
{q^2-m_\omega^2} A^{\nu\alpha\beta\gamma},\ \ \
	G^\mu=\gamma^\mu F_{\omega NN}(q_1^2),\nonumber\\
 A^{\nu\alpha\beta\gamma}&=&
	~T^{\alpha\nu}_{k_1,q_1}F_{YJ/\psi\omega}(q_1^2)
	\frac{1}{q_2^2-M_Y^2+iM_Y\Gamma}
	~T'^{\beta\gamma}_{k_2,k_3}F_{YJ/\psi\omega}(q_2^2)\nonumber.\end{eqnarray}

\begin{center}
\begin{figure}[htb]
\begin{tabular}{cc}
\scalebox{0.7}{\includegraphics{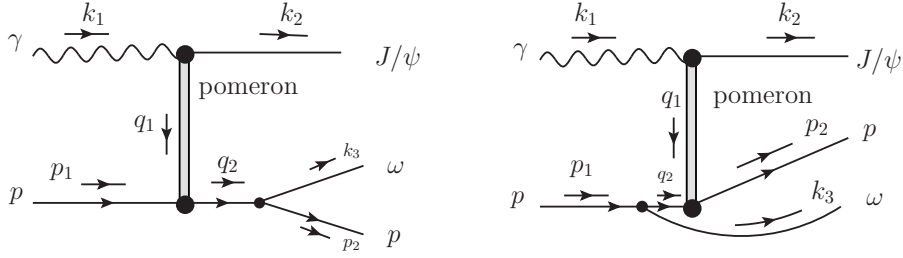}}
\end{tabular}
\caption{The $\gamma p\to J/\psi\omega p$ process via the pomeron exchange.\label{Fig: P}}
\end{figure}
\end{center}

Since the pomeron mediates the long-range interaction between a confined
quark and a nucleon, thus $\gamma p\to J/\psi\omega p$ occurs via the pomeron exchange, which is the main background contribution to $\gamma p\to Y(3940)p\to J/\psi\omega p$.
We adopt the formulas in Refs. ~\cite{Donnachie:1987pu,Pichowsky:1996jx,Liu:2008qx} to
describe the pomeron exchange process shown in Fig. \ref{Fig: P}.

The pomeron-nucleon coupling is determined by the vertex
\begin{eqnarray}
F^\mu(t)&=&\frac{3\beta_0(4M_N^2-2.8t)}{(4M_N^2-t)(1-t/0.7)^2}
\gamma^\mu=F(t)\gamma^\mu ,\nonumber
\end{eqnarray}
where $t=q^2_1$ is the exchanged pomeron momentum squared. $\beta_0^2=4$ GeV$^2$ denotes the coupling
constant between a single pomeron and a light constituent quark.

For the $\gamma V{\cal P}$ vertex with on-shell approximation for restoring the gauge invariance, one has
\begin{eqnarray}
	V_{\gamma V\cal P}&=&\frac{2\beta_c~4\mu_0^2}{(M_V^2-t)(2\mu_0^2+M_V^2-t)}
	T_{\alpha\nu\beta}\epsilon_V^\beta\epsilon_\gamma^\alpha {\cal P}^\nu=V(t)
	T_{\alpha\nu\beta}\epsilon_V^\beta\epsilon_\gamma^\alpha {\cal P}^\nu,\\
	T_{\alpha\nu\beta}&=&(k_\gamma+k_V)_\nu g_{\alpha\beta}-2k_{\gamma\beta} g_{\nu \alpha},
\end{eqnarray}
where $\beta_c^2=0.8$ GeV$^2$ and $\mu_0=1.05$ GeV.

The amplitudes for the $s-$ and $u-$channel, corresponding to the diagrams on the left and on the right in Fig. \ref{Fig: P} respectively, can be written as
\begin{eqnarray}
	{\cal T}_s^{\cal P}&=&G_{cc,ff}T_{\alpha\nu\beta}
	\bar{u}(p_2)\gamma_\xi (\rlap\slash q_2+m_N)\gamma^\nu u(p_1)
	\epsilon_\gamma^\alpha\epsilon_V^\beta \epsilon_\omega^\xi,\\
	{\cal T}_u^{\cal P}&=&G_{cc,ff}T_{\alpha\nu\beta}
	\bar{u}(p_2)\gamma^\nu(\rlap\slash q_2+m_N)\gamma_\xi u(p_1)
	\epsilon_\gamma^\alpha\epsilon_V^\beta \epsilon_\omega^\xi
\end{eqnarray}
with $G_{cc,ff}=g_{\omega NN}F_{\omega NN}(q_2^2) F(t)V(t){\cal
G}_P(s,t)/(q_2^2-m_N^2)$, where ${\cal G}_P(s,t)$ is related to the pomeron
trajectory $\alpha(t)=1+\epsilon+\alpha' t$ via ${\cal
G}_P(s,t)=-i(\alpha's)^{\alpha(t)-1}$ with $\alpha'=0.25$ GeV$^{-2}$ and
$\epsilon=0.08$. $F_{\omega NN}(q_2^2)=({\Lambda^2_{\omega
NN}-m_N^2})/({\Lambda^2_{\omega NN}-q_2^2})$.

\begin{figure}[h!]
\includegraphics[bb=200 550 380 770 ,scale=1.2]{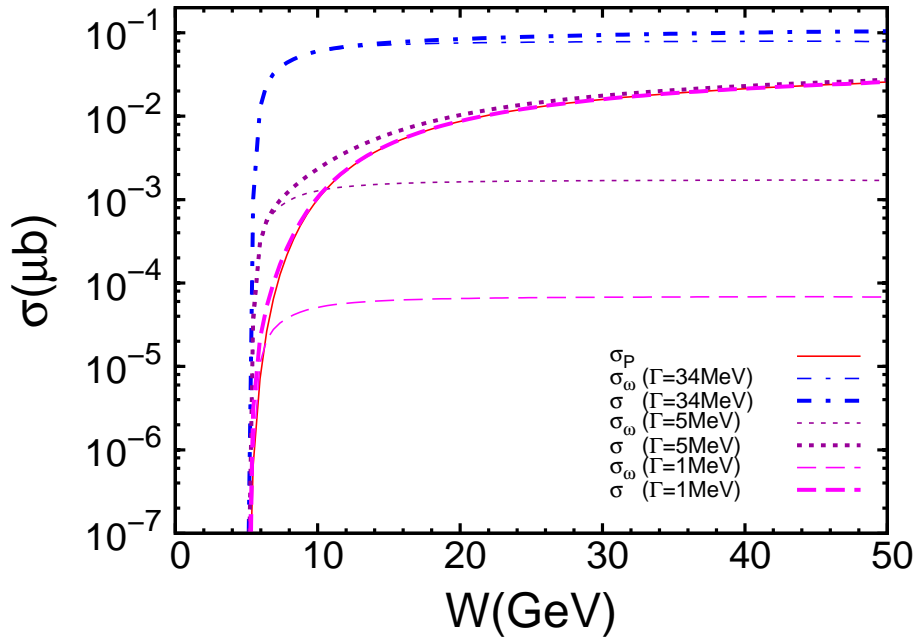}
\caption{
The dependence of the total cross sections of $\gamma p\to J/\psi\omega p$ on
$-q^2$. Here, $\sigma_P$ denotes the total cross section of $\gamma p\to J/\psi\omega
p$ via the pomeron exchange. $\sigma$ is the total cross section of $\gamma p\to
J/\psi\omega p$ via both pomeron and $\omega$ exchanges shown in Figs.
\ref{Fig:gppop} and \ref{Fig: P}. $\sigma_\omega$ means the total cross section
of $\gamma p\to J/\psi\omega p$ through exchanging $\omega$, which is related
to $Y(3940)$ production directly.  Since $\sigma_\omega$ is proportional to $g^4_{YJ/\psi \omega}$ determined by the decay width of $Y(3940)\to
J/\psi\omega$, we give the variation of $\sigma_\omega$ to $W$ with
several typical values of $g_{YJ/\psi \omega}$, which correspond to
$\Gamma_{YJ/\psi\omega}=1,\,5,\, 34$ MeV.
\label{Fig:gNpsiomNTCS} }
\end{figure}

With the transition amplitude ${\mathcal T}$, we obtain the total cross section of the $\gamma p\to J/\psi\omega p$ process
\begin{eqnarray}
	d\sigma&=&\frac{m_N}{2 k_1\cdot p_1}|{\cal T}|^2
	\frac{d^3k_2}{(2\pi)^3}\frac{1}{2k^0_{2}}\frac{d^3k_{3}}{(2\pi)^3}\frac{1}{2k^0_{3}}
	\frac{d^3p_{2}}{(2\pi)^3}\frac{m_N}{p^0_{2}}
	(2\pi)^4\delta^4(k_1+p_1-k_2-k_3-p_2).
\end{eqnarray}
The numeral results are obtained by the FOWL program, which is presented in Fig. \ref{Fig:gNpsiomNTCS}.

The line shape of the total cross section of $\gamma p\to J/\psi \omega p$ via
$\omega$ exchange (see Fig. \ref{Fig:gNpsiomNTCS}) is similar to that of
$\gamma p\to Y(3940) p$ [see Fig. \ref{Fig:gNYNTCS} (b)], which goes up very
rapidly near the threshold and becomes flat with increasing $W$.  The line shape
of the total cross section of $\gamma p\to J/\psi \omega p$ through the pomeron
exchange is monotonically increasing.

\begin{figure}[htb]
\includegraphics[bb=0 160 580 660 ,scale=0.26]{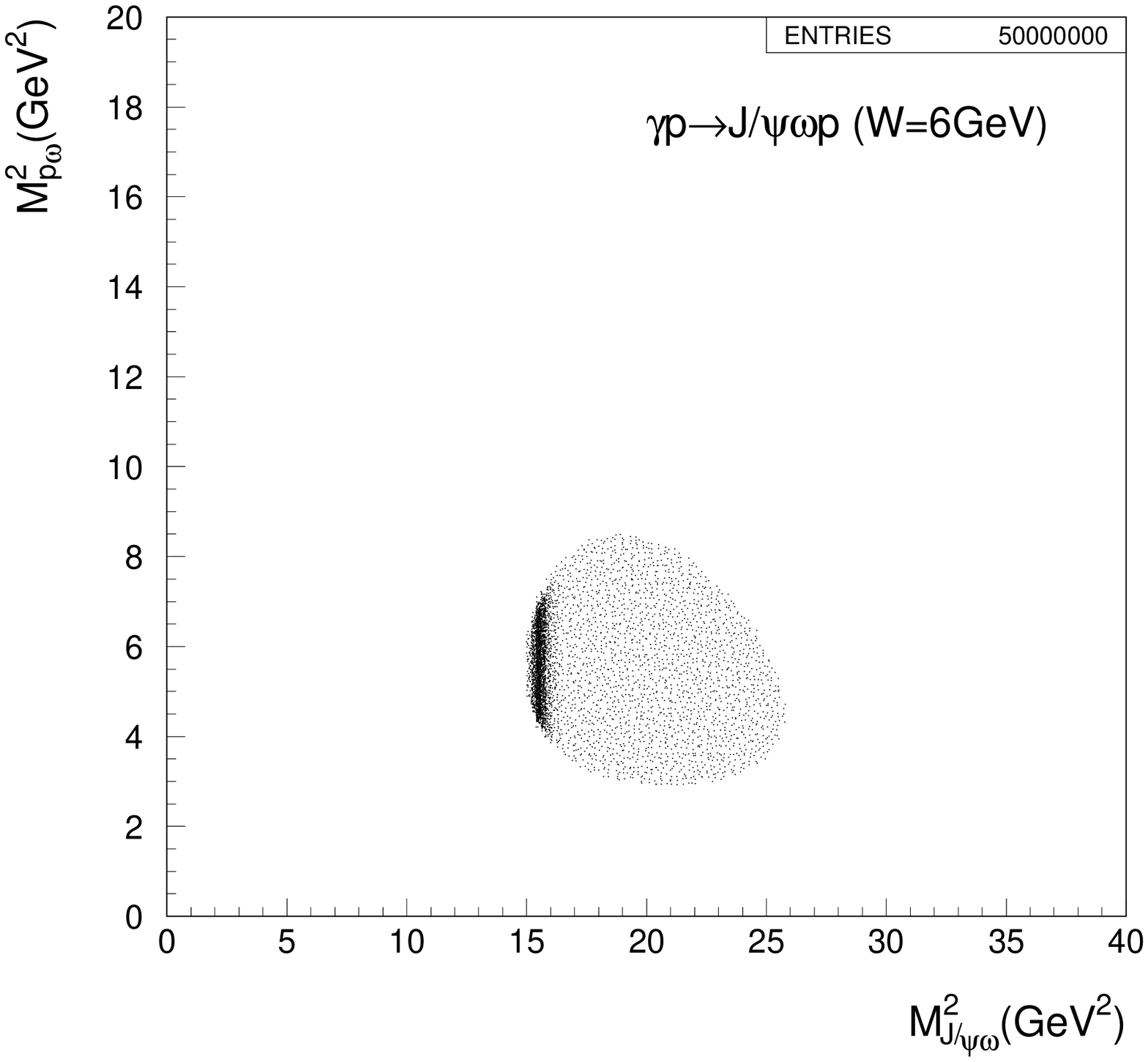}
\includegraphics[bb=0 160 580 660 ,scale=0.26]{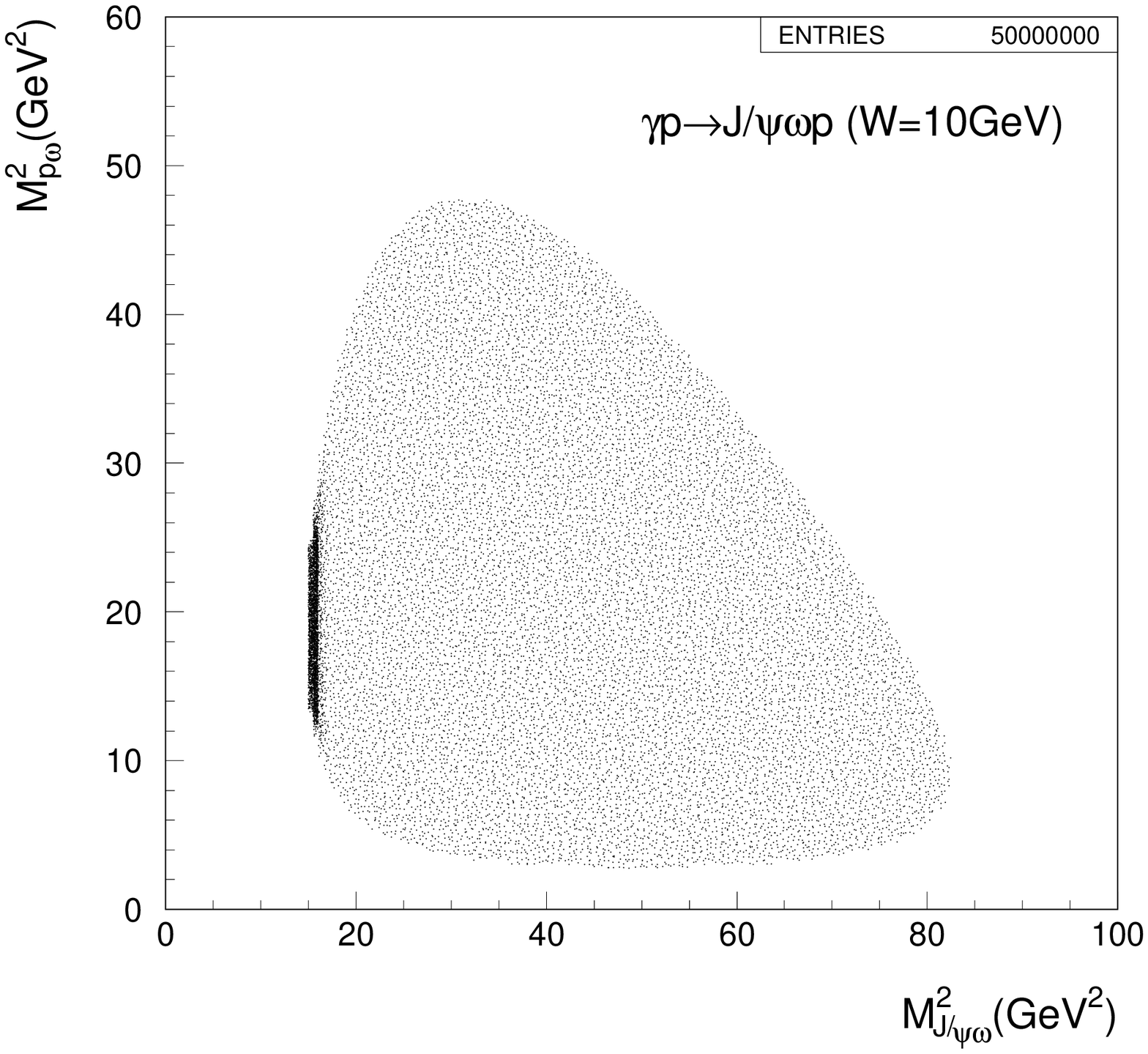}
\includegraphics[bb=0 160 580 660 ,scale=0.26]{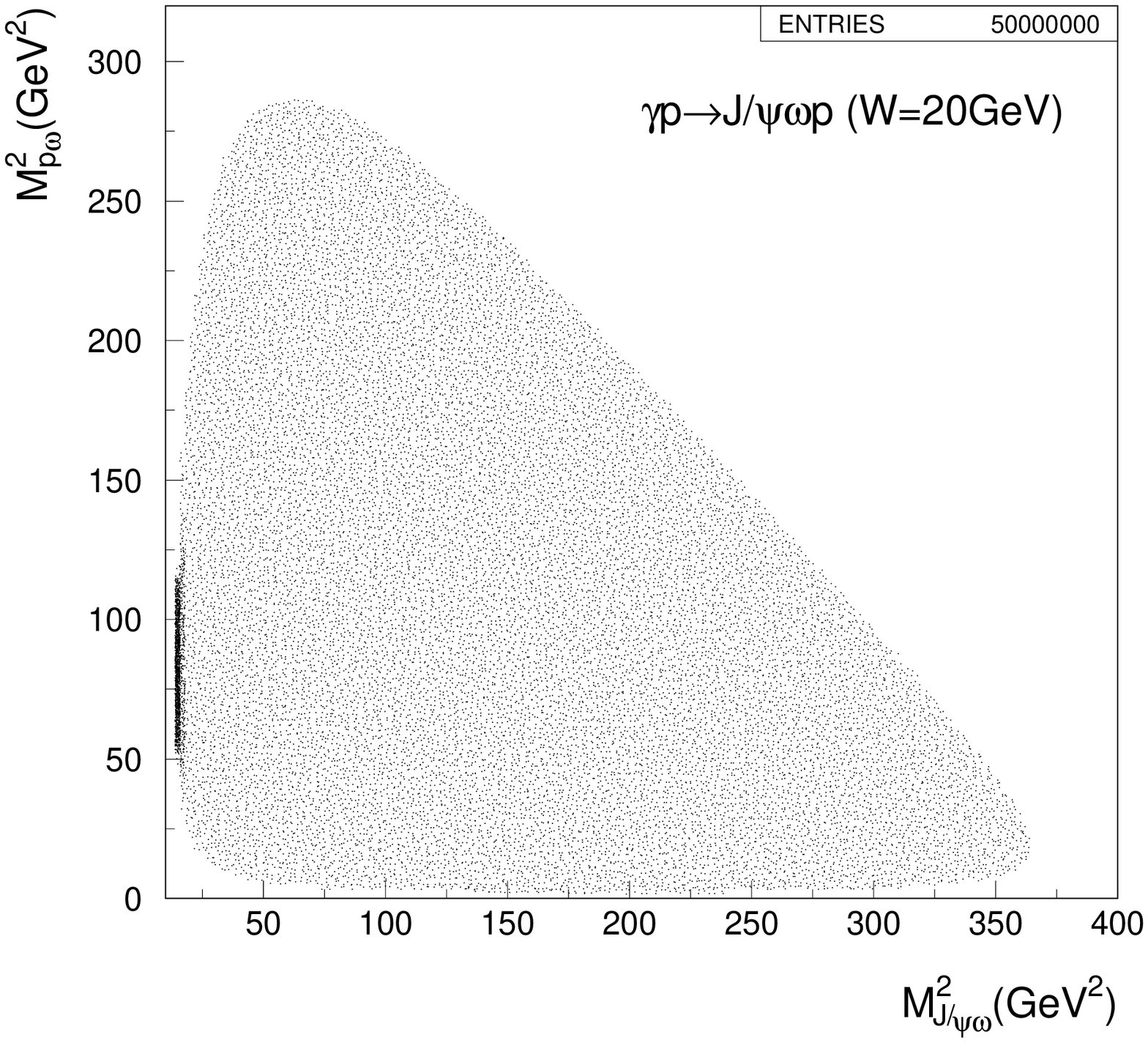}
\includegraphics[bb=0 160 580 660 ,scale=0.26]{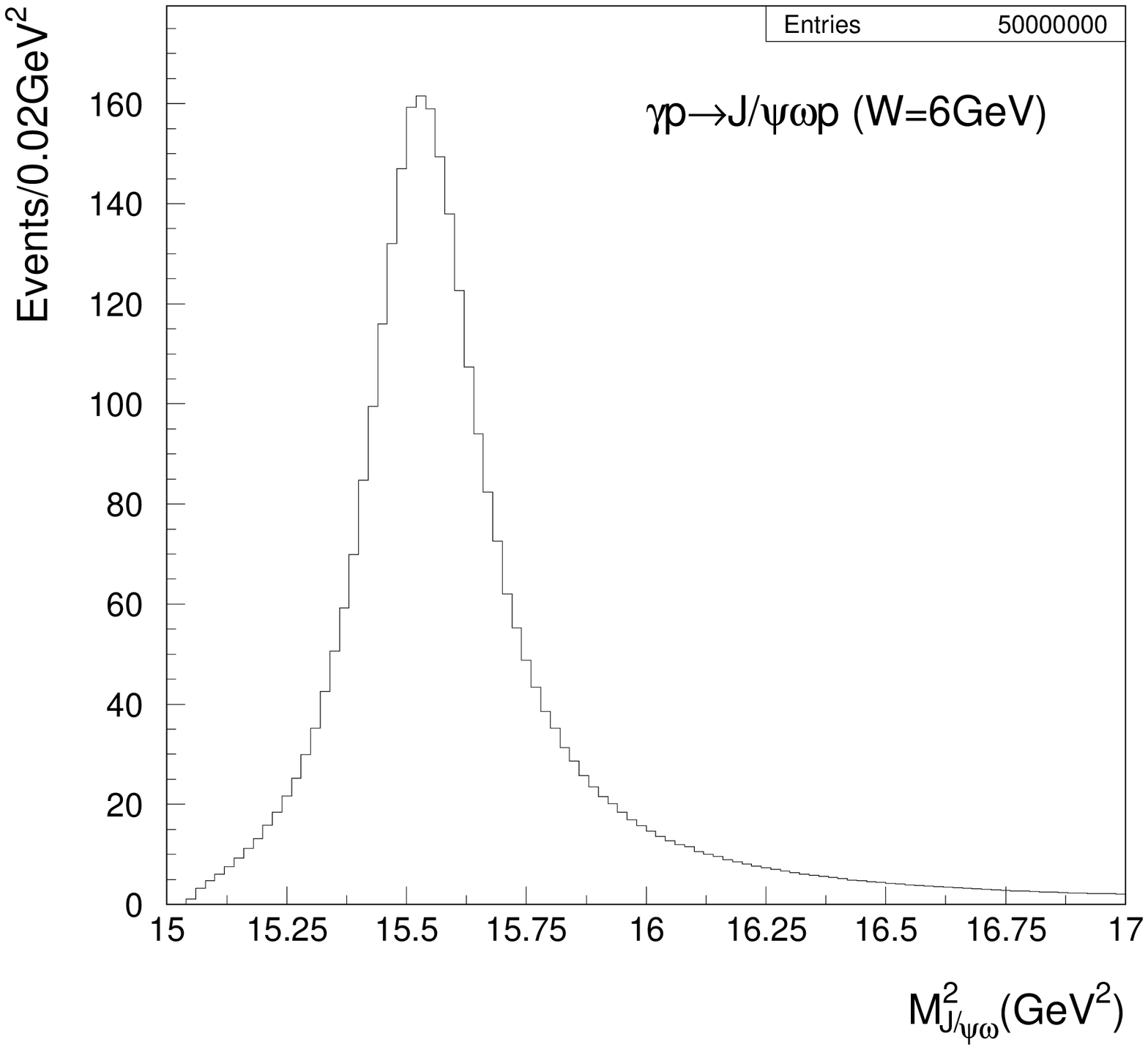}
\includegraphics[bb=0 160 580 660 ,scale=0.26]{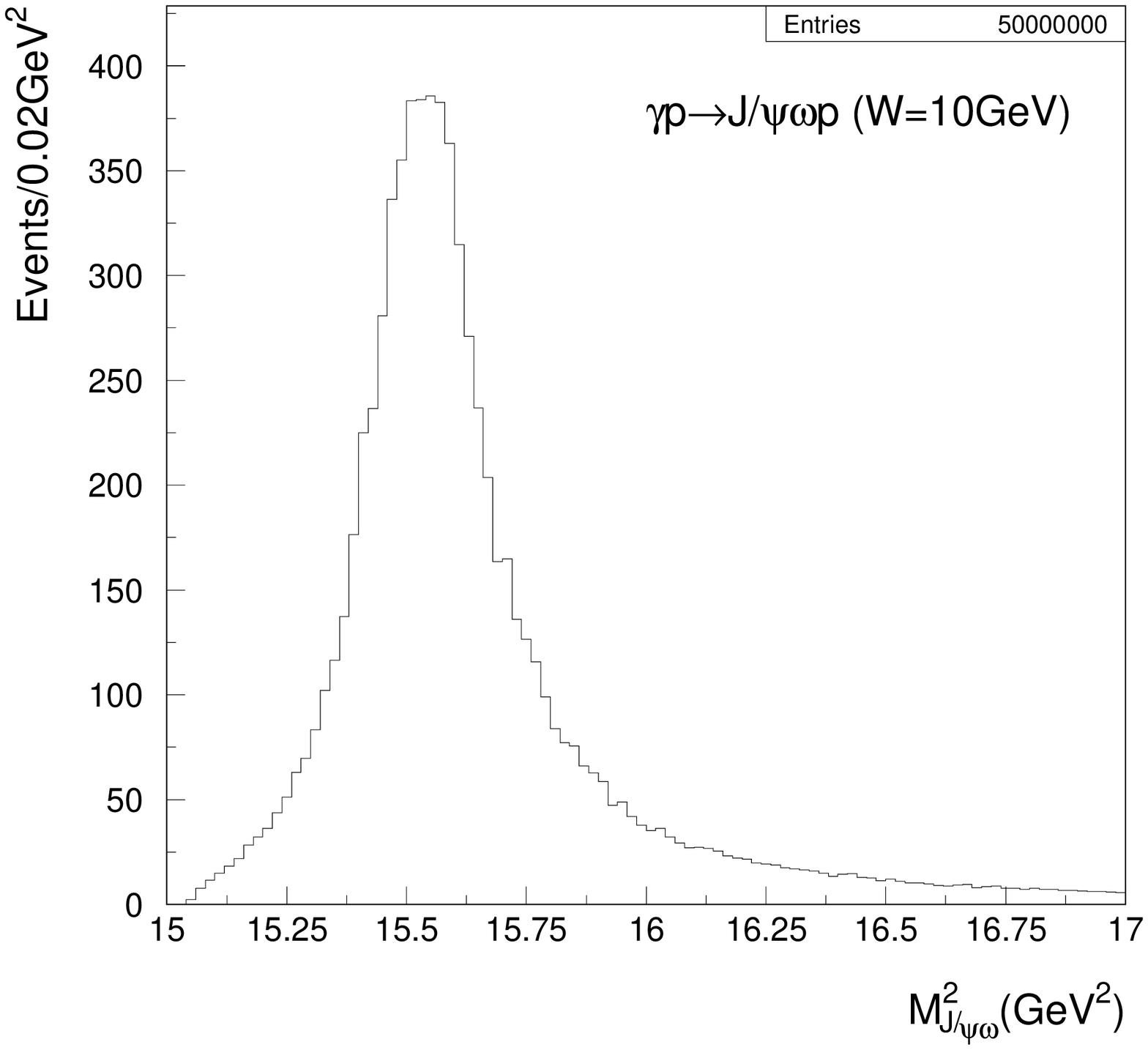}
\includegraphics[bb=0 160 580 660 ,scale=0.26]{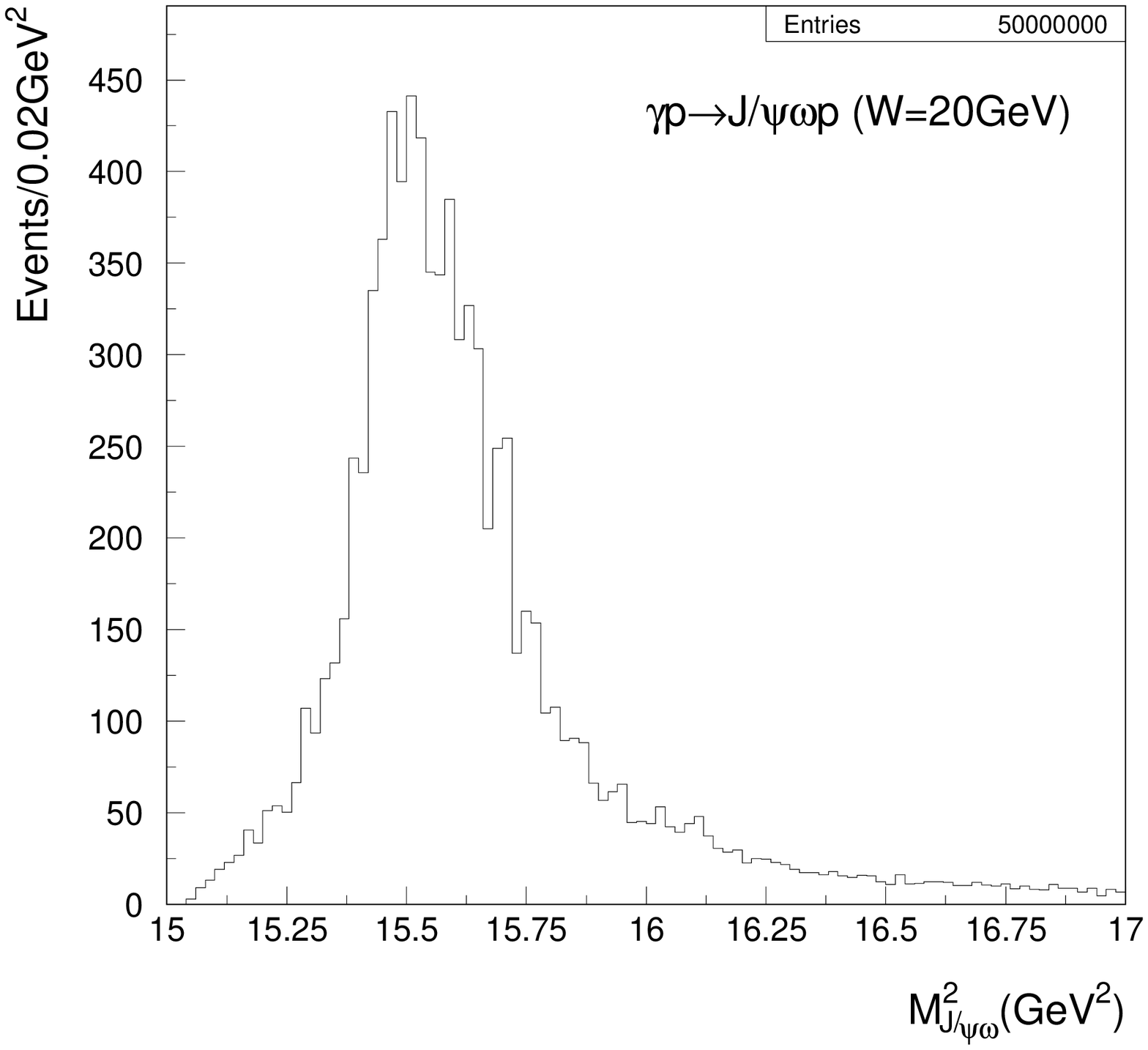}
\caption{The Dalitz plot (above) and the $J/\psi\omega$ invariant mass spectrum (bottom) for $\gamma p\to J/\psi\omega p$ process with the scalar $Y(3940)$ production. Here, the numerical result corresponds to the upper limit of $\Gamma_{Y J/\psi\omega}$.  \label{Fig:Dalitz plot}}
\end{figure}

By the Dalitz plot, we can identify $Y(3940)$ by analyzing the invariant mass
spectrum of $J/\psi\omega$.
In Fig. \ref{Fig:Dalitz plot}, the Dalitz plot and the corresponding
$J/\psi\omega$ invariant mass spectrum with several typical values of $W$ are
presented, where the numerical result is obtained under the upper limit of
$\Gamma_{Y J/\psi\omega}$.  As shown in Fig. \ref{Fig:Dalitz plot}, an
explicit band corresponding to $Y(3940)$ appears even at the higher energy region
of $W$, which indicates there exists a wide energy window of $W$ to identify
$Y(3940)$ in experiments. By analyzing the invariant mass spectrum of
$J/\psi \omega$, one finds that the number of events of $Y(3940)$ is up to
450/0.02GeV$^2$ in $50\times 10^6$ collisions of $\gamma p$.

\begin{figure}[h!]
\includegraphics[bb=0 160 580 660 ,scale=0.4]{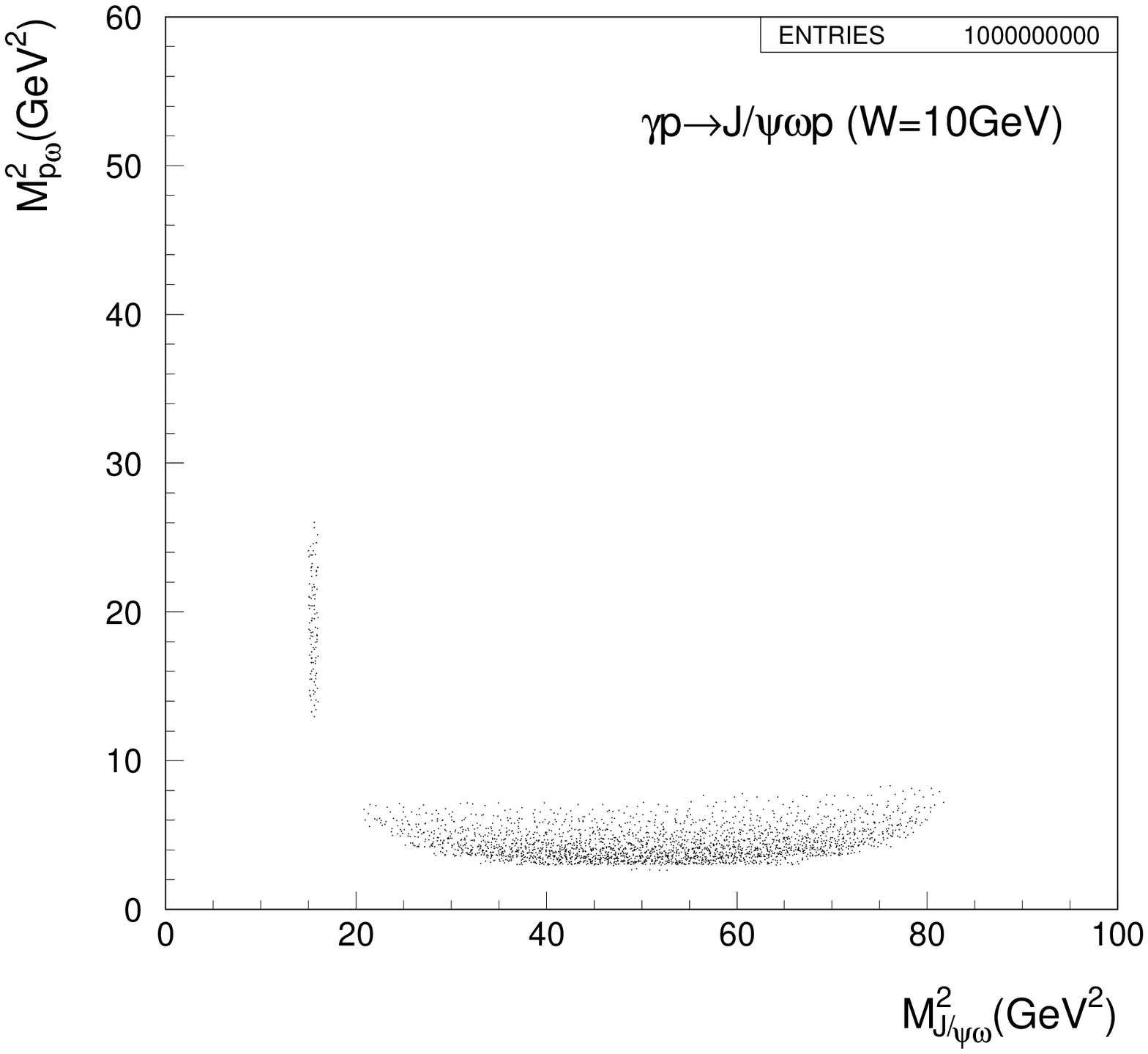}
\includegraphics[bb=0 160 580 660 ,scale=0.4]{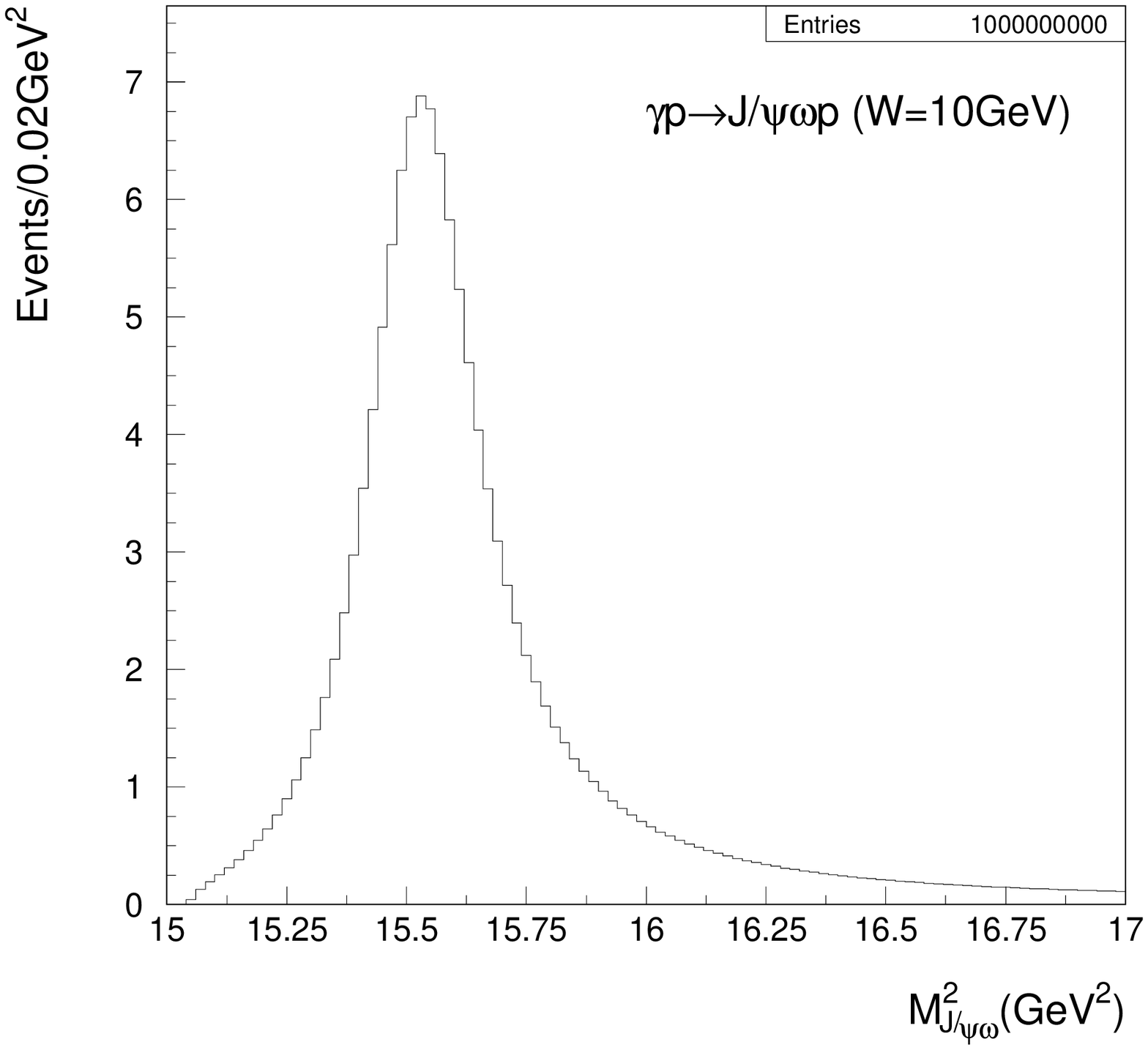}
\caption{The Dalitz plot (left) and the $J/\psi\omega$ invariant mass spectrum (right) for the $\gamma p\to J/\psi\omega p$ with the scalar $Y(3940)$ if taking the lower limit of $g_{Y J/\psi\omega}$ as
the parameter input.  \label{Fig:Dalitz plot0}}
\end{figure}

If a smaller decay width of $Y(3940)\to J/\psi\omega$ is adopted, the number of
events of $Y(3940)$ is reduced. We can expect that the background
contribution from the pomeron exchange appears in the Dalitz plot.  Here, with the
numerical result corresponding to the upper limit of decay width
$\Gamma_{YJ/\psi\omega}$ as an example, the Dalitz plot of $\gamma p\to
J/\psi\omega p$ and the invariant mass spectrum of $J/\psi\omega$ are
illustrated in Fig.~\ref{Fig:Dalitz plot0}.  The number of events of $Y(3940)$
decreases to about 10/0.02GeV$^2$ when taking $10^9$ collisions of
$\gamma p$. There exists an obvious pomeron exchange contribution, a band in
the bottom of the Dalitz plot. However, the bands respectively corresponding to $Y(3940)$
 and the background contribution from the pomeron exchange do not
interfere with each other. Thus, we still can distinguish the signal of
$Y(3940)$ from the background in the Dalitz plot. We need to emphasize that the numerical results are not sensitive to the cutoffs
$\Lambda_Y/\Lambda_{\omega}$.

\section{conclusion and discussion}\label{Sec:Conclu}

In this work, we study the possibility to search for $Y(3940)$ by the
photoproduction process. Since $Y(3940)$ was observed in the invariant mass spectrum of $J/\psi\omega$, $\gamma p\to Y(3940)p$ by exchanging the $\omega$
meson is the main channel to produce $Y(3940)$. Our numerical result shows that the upper (lower) limit of the total cross section for the $\gamma p\to Y(3940)p$ is on the order of 1 $nb$ (0.1 $\mu b$),
which is comparable with the cross section of $J/\psi$, $\psi(2S)$ and $\Upsilon$ photoproduction in the HERA experiment.

Additionally we further carry out the background analysis relevant to the
production of $Y(3940)$, where $\gamma p\to J/\psi\omega p$ occurs via the pomeron
exchange.  By the Dalitz plot, we find that the $Y(3940)$ signal can be
distinguished from the background clearly. The result of the invariant mass
spectrum of $J/\psi\omega$ indicates the lower limit of the number of events of
$Y(3940)$ can reach up to 10/0.02GeV$^2$ for $1\times 10^9$ collisions
of $\gamma p$, which shows that there exists a potential to find $Y(3940)$
 by meson photoproduction process. Since the calculation in this work
is relevant to the decay width of ${Y(3940)\to J/\psi\omega}$, to some extent
we encourage our experimental colleagues to carry out the measurement of the decay
width of ${Y(3940)\to J/\psi\omega}$, which will be helpful to make further
predictions of $Y(3940)$ production by the photoproduction process.

As we all know, most of the charmonium-like states $X$, $Y$, $Z$ are observed by $B$ meson decay. Thus, searching for these $X$, $Y$, $Z$ states by other
processes will be helpful to establish them. Besides studying the production of $Y(3940)$ and $Z(4430)$ in the
meson photoproduction process, exploring the production of remaining $X$, $Y$, $Z$ states by meson photoproduction will be an interesting topic. The experimental search for the charmonium-like states $X$, $Y$, $Z$
is encouraged, especially for the HERA experiment.

\section*{Acknowledgements}

We are grateful to Dr. Xu Cao for communication of the FOWL program. This project is supported by the National Natural Science Foundation of China under Grants No. 10705001 and No. 10905077 and
the Foundation for the Author of National Excellent Doctoral Dissertation of P.R. China (FANEDD).


\begin{thebibliography}{98}
\bibitem{Abe:2004zs}
  K.~Abe {\it et al.}  [Belle Collaboration],
  Phys.\ Rev.\ Lett.\  {\bf 94}, 182002 (2005)
  [arXiv:hep-ex/0408126].

\bibitem{Aubert:2007vj}
  B.~Aubert {\it et al.}  [BaBar Collaboration],
  Phys.\ Rev.\ Lett.\  {\bf 101}, 082001 (2008)
  [arXiv:0711.2047 [hep-ex]].

\bibitem{Zhu:2007wz}
  S.~L.~Zhu,
  Int.\ J.\ Mod.\ Phys.\  E {\bf 17}, 283 (2008)
  [arXiv:hep-ph/0703225].

\bibitem{Godfrey:2008nc}
  S.~Godfrey and S.~L.~Olsen,
  Ann.\ Rev.\ Nucl.\ Part.\ Sci.\  {\bf 58}, 51 (2008)
  [arXiv:0801.3867 [hep-ph]].

\bibitem{Aaltonen:2009tz}
  T.~Aaltonen {\it et al.}  [CDF Collaboration],
  Phys.\ Rev.\ Lett.\  {\bf 102}, 242002 (2009)
  [arXiv:0903.2229 [hep-ex]].


\bibitem{Liu:2009ei}
  X.~Liu and S.~L.~Zhu,
  Phys.\ Rev.\  D {\bf 80}, 017502 (2009)
  [arXiv:0903.2529 [hep-ph]];
  X.~Liu and H.~W.~Ke,
  Phys.\ Rev.\  D {\bf 80}, 034009 (2009)
  [arXiv:0907.1349 [hep-ph]];
  X.~Liu,
  Phys.\ Lett.\  B {\bf 680}, 137 (2009)
  [arXiv:0904.0136 [hep-ph]].


\bibitem{Branz:2009yt}
  T.~Branz, T.~Gutsche and V.~E.~Lyubovitskij,
  arXiv:0903.5424 [hep-ph].

\bibitem{:2007wga}
  S.~K.~Choi {\it et al.}  [BELLE Collaboration],
  Phys.\ Rev.\ Lett.\  {\bf 100}, 142001 (2008)
  [arXiv:0708.1790 [hep-ex]].


\bibitem{Liu:2008qx}
  X.~H.~Liu, Q.~Zhao and F.~E.~Close,
  Phys.\ Rev.\  D {\bf 77}, 094005 (2008)
  [arXiv:0802.2648 [hep-ph]].

\bibitem{Ke:2008kf}
  H.~W.~Ke and X.~Liu,
  Eur.\ Phys.\ J.\  C {\bf 58}, 217 (2008)
  [arXiv:0806.0998 [hep-ph]].

\bibitem{Chekanov:2002xi}
  S.~Chekanov {\it et al.}  [ZEUS Collaboration],
  Eur.\ Phys.\ J.\  C {\bf 24}, 345 (2002)
  [arXiv:hep-ex/0201043].

\bibitem{Chekanov:2002at}
  S.~Chekanov {\it et al.}  [ZEUS Collaboration],
  Eur.\ Phys.\ J.\  C {\bf 27}, 173 (2003)
  [arXiv:hep-ex/0211011].


\bibitem{Chekanov:2004mw}
  S.~Chekanov {\it et al.}  [ZEUS Collaboration],
  Nucl.\ Phys.\  B {\bf 695}, 3 (2004)
  [arXiv:hep-ex/0404008].

\bibitem{Black:2002ek}
  D.~Black, M.~Harada and J.~Schechter,
  Phys.\ Rev.\ Lett.\  {\bf 88}, 181603 (2002)
  [arXiv:hep-ph/0202069].



\bibitem{Oh:2003aw}
  Y.~s.~Oh and T.~S.~H.~Lee,
  Phys.\ Rev.\  C {\bf 69}, 025201 (2004)
  [arXiv:nucl-th/0306033].

\bibitem{Renner:1971sj}
  B.~Renner,
  Nucl.\ Phys.\  B {\bf 30}, 634 (1971).


\bibitem{Lansberg:2009xh}
  J.~P.~Lansberg and T.~N.~Pham,
  Phys.\ Rev.\  D {\bf 79}, 094016 (2009)
  [arXiv:0903.1562 [hep-ph]].



\bibitem{Amsler:2008zzb}
  C.~Amsler {\it et al.}  [Particle Data Group],
  Phys.\ Lett.\  B {\bf 667}, 1 (2008).

\bibitem{Gasparyan:2003fp}
  A.~M.~Gasparyan, J.~Haidenbauer, C.~Hanhart and J.~Speth,
  Phys.\ Rev.\  C {\bf 68}, 045207 (2003)
  [arXiv:nucl-th/0307072].


\bibitem{Chekanov:2009xb}
  S. Chekanov {\it et al.} [ZEUS Collaboration],
  arXiv:0903.4205 [hep-ex].

\bibitem{Breitweg:1998ki}
  J.~Breitweg {\it et al.}  [ZEUS Collaboration],
  Phys.\ Lett.\  B {\bf 437}, 432 (1998)
  [arXiv:hep-ex/9807020].



\bibitem{Donnachie:1987pu}
  A.~Donnachie and P.~V.~Landshoff,
  Phys.\ Lett.\  B {\bf 185}, 403 (1987).



\bibitem{Pichowsky:1996jx}
  M.~A.~Pichowsky and T.~S.~H.~Lee,
  Phys.\ Lett.\  B {\bf 379}, 1 (1996)
  [arXiv:nucl-th/9601032].






\end{thebibliography}
\end{document}